\documentclass[12pt]{article}

\usepackage[utf8]{inputenc}
\usepackage[T1]{fontenc}
\usepackage[margin=1in]{geometry}
\usepackage{amsmath,amssymb,amsthm}
\usepackage{booktabs}
\usepackage{array}
\usepackage{hyperref}
\usepackage{url}
\usepackage{graphicx}
\usepackage{xcolor}
\usepackage[numbers,sort&compress]{natbib}
\usepackage{authblk}
\usepackage{setspace}
\usepackage{enumitem}
\usepackage{microtype}
\usepackage{pgfplots}
\pgfplotsset{compat=1.18}
\usepackage{siunitx}
\usepackage{tabularx, booktabs, array}
\usepackage{listings}
\usepackage{caption}

\lstdefinelanguage{Rust}{
	keywords={enum, match, fn, pub, struct, impl, let, mut, return, use,
		mod, type, where, Self, self, true, false},
	keywordstyle=\bfseries\color{blue!70!black},
	comment=[l]{//},
	commentstyle=\itshape\color{gray},
	stringstyle=\color{orange!80!black},
	basicstyle=\ttfamily\small,
	breaklines=true,
	frame=single,
	rulecolor=\color{gray!40},
	backgroundcolor=\color{gray!5},
	captionpos=b,
	numbers=left,
	numberstyle=\tiny\color{gray},
	xleftmargin=2em,
}

\newtheorem{definition}{Definition}[section]
\newtheorem{property}{Design Property}[section]
\newtheorem{remark}{Remark}[section]

\newcommand{\Blocklace}{\mathcal{B}}
\newcommand{\Draw}{\mathcal{D}_{\text{raw}}}
\newcommand{\Dpub}{\mathcal{D}_{\text{pub}}}
\newcommand{\SA}{\textsc{sa}}
\newcommand{\FSTP}{\textsc{fstp}}
\newcommand{\GII}{\textsc{gii}}
\newcommand{\CII}{\textsc{cii}}
\newcommand{\Fr}{\mathrm{Fr}}

\hypersetup{colorlinks=true,linkcolor=black,citecolor=black,urlcolor=blue}

\title{
	\textbf{Federated Sovereign Transport Protocol (FSTP):}\\
	\textbf{Verifiable Coordination Without Disclosure}
}

\author[1]{Ram\'on Soto C.}
\author[2]{Liz Soto}
\affil[1]{Department of Accounting, University of Sonora,
          Hermosillo, Sonora, Mexico \\ \nolinkurl{ramon.soto@unison.mx}}
\affil[2]{Department of Mathematics, University of Sonora,
	Hermosillo, Sonora, Mexico}

\date{}

\begin{document}
	\maketitle
	
	\begin{abstract}
		This paper introduces the Federated Sovereign Transport Protocol (\FSTP{}),
		a synchronization boundary and transport layer for federated networks
		in which nodes have heterogeneous privacy requirements. Existing
		federation protocols leave data confinement to operator policy: they
		define message formats and delivery semantics but impose no structural
		constraint on what a conforming server may emit. \FSTP{} addresses this
		gap by making data confinement a property of the protocol itself.
		
		The central mechanism is a synchronization agent whose output type set
		is formally closed. Raw internal data cannot appear in any federation
		message because the constraint is enforced by the Rust type system at
		compile time, not by a runtime check. A received artifact carries a
		usage scope: a conforming agent refuses to re-emit it toward a third
		node unless that scope, or a fresh governance grant, authorizes the
		recipient. A contextual identity model
		derives a separate, unlinkable identifier for each federation
		relationship, preventing cross-context correlation structurally. A
		Blocklace-based event substrate~\cite{shapiro2023grassroots,keidar2023cordial}
		provides tamper-evident, partially ordered logging with synchronization
		cost proportional to the symmetric difference between node states,
		and supports data erasure without breaking the hash chain.
		
		The result is \textbf{proof without exposure}: a federation participant
		can verify that a process occurred, that a credential is authentic, and
		that an outcome is uncorrupted without accessing the internal data that
		produced these artifacts. \FSTP{} is developed as the inter-node
		transport layer of Velyzor, a governance platform for institutions with
		demanding confidentiality requirements. The specification and reference
		implementation are released as open-source infrastructure under
		Apache~2.0; source code and figures accompany this paper.
		
	\end{abstract}

	\section{Introduction}
	\label{sec:intro}
	
	\subsection{The Gap in Existing Federation Protocols}
	
	Federated social platforms have demonstrated that large-scale digital
	coordination does not require a central authority. Protocols such as
	ActivityPub~\cite{w3c2018activitypub} allow independently operated
	servers to form a coherent network while each retains administrative
	autonomy. This architecture rests on an implicit assumption: that all
	nodes are willing to expose their internal activity to the network.
	Messages may pass through intermediate nodes, user identities are
	globally resolvable, and participation is visible to any node that
	receives a copy of the relevant events.
	
	For many participants this exposure is acceptable. For entities
	operating under data protection regulations, sector-specific
	confidentiality obligations, or strong institutional privacy
	positions---trade unions, political parties, professional associations,
	regulated firms---it is not. These entities need to interact with a
	broader network while retaining exclusive custody of their internal
	data: membership records, deliberation histories, identity attributes,
	and operational information.
	
	Existing federation protocols do not address this requirement at the
	protocol level. They define message formats and delivery semantics;
	they do not constrain what a conforming server may emit, nor guarantee
	that internal data stays internal. Where trust in data confinement
	exists, it rests on operator policy rather than on protocol structure.
	ActivityPub provides visibility modifiers such as ``unlisted'' posts,
	but these are application-layer conventions: a conforming peer node
	may receive and redistribute content regardless of these markers
	because the protocol imposes no structural barrier to doing so. For
	institutional data---deliberation records, member votes, identity
	attributes---a convention-level restriction is insufficient; the
	guarantee must be structural and hold against a conforming but
	adversarial peer.
	
	\subsection{FSTP: Confinement as a Protocol Property}
	
	\FSTP{} addresses this gap by making data confinement a structural
	property of the protocol rather than a matter of operational policy.
	The main mechanism is a \textbf{synchronization agent} (\SA{}) that
	sits at the node boundary as the exclusive interface between the
	node's internal data store and the federation network. The \SA{}
	accepts only a formally closed set of typed cryptographic objects as
	output, and this constraint is enforced by the type system of the
	reference implementation as a compile-time guarantee, not a runtime
	check.
	
	The reference implementation is written in Rust. The confinement
	guarantee is realized through Rust's \texttt{enum} types with
	exhaustive pattern matching: the compiler statically rejects any code
	path that attempts to place a value of an internal data type into a
	federation message. Section~\ref{sec:typesystem} presents the type
	model and an illustrative code fragment.
	
	Two additional mechanisms complete the protocol. A
	\textbf{contextual identity model} allows users to present verifiable
	credentials across node boundaries under relationship-specific
	identities that cannot be linked to their global identity or to
	identities in other relationships. A \textbf{Blocklace
		substrate}~\cite{shapiro2023grassroots,keidar2023cordial} provides
	tamper-evident, partially ordered event logging whose synchronization
	cost is proportional to the difference between node states, and whose
	structure is compatible with data erasure obligations without loss of
	hash-chain integrity.
	
	Together these three mechanisms realize \textbf{proof without
		exposure}: a federation participant can verify that a process
	occurred, that a credential is authentic, and that an outcome is
	uncorrupted---without accessing the internal data that produced these
	artifacts.
	
	\subsection{Motivating Context}
	\label{sec:motivation}
	
	Distributed institutions face a coordination problem that existing
	transport layers cannot solve. Methods for iterative group decision-making
	in structured social networks, such as the fuzzy aggregation
	approach of~\cite{soto2016mqdm}, produce collective outcomes that are
	certifiably legitimate precisely because they preserve the integrity
	of each participant's deliberative process. That integrity depends on
	a transport layer capable of certifying outcomes across organizational
	boundaries \emph{without} exposing the underlying deliberative record.
	\FSTP{} is designed to provide exactly this link: the hash of a
	certified collective decision crosses the node boundary as verifiable
	proof, while the deliberation that produced it remains under
	institutional custody.
	
	\FSTP{} is not a purely academic proposal. It serves as the inter-node
	transport layer of Velyzor, a governance platform for institutions with
	demanding confidentiality requirements, and is designed for
	operational deployability under real connectivity conditions. Data
	sovereignty in \FSTP{} is a structural consequence of deployment, not
	a feature requiring conscious activation.
	
	\subsection{Paper Organization}
	
	Section~\ref{sec:overview} presents the system model, trust
	assumptions, and a comparison with related approaches.
	Section~\ref{sec:design} describes the three protocol primitives in
	detail. Section~\ref{sec:security} discusses security and privacy
	properties and gives a protocol-level privacy audit.
	Section~\ref{sec:scenarios} develops a coordination case study and
	additional coordination patterns. Section~\ref{sec:discussion}
	addresses deployment considerations, regulatory compatibility, and
	licensing. Section~\ref{sec:benchmarks} reports empirical
	synchronization benchmarks for the reference implementation.
	Section~\ref{sec:conclusions} concludes.

	\section{System Model and Related Work}
	\label{sec:overview}
	
	\subsection{Architecture and Information Partition}
	
\begin{figure}[h]
		\centering
		\includegraphics[width=0.75\textwidth]{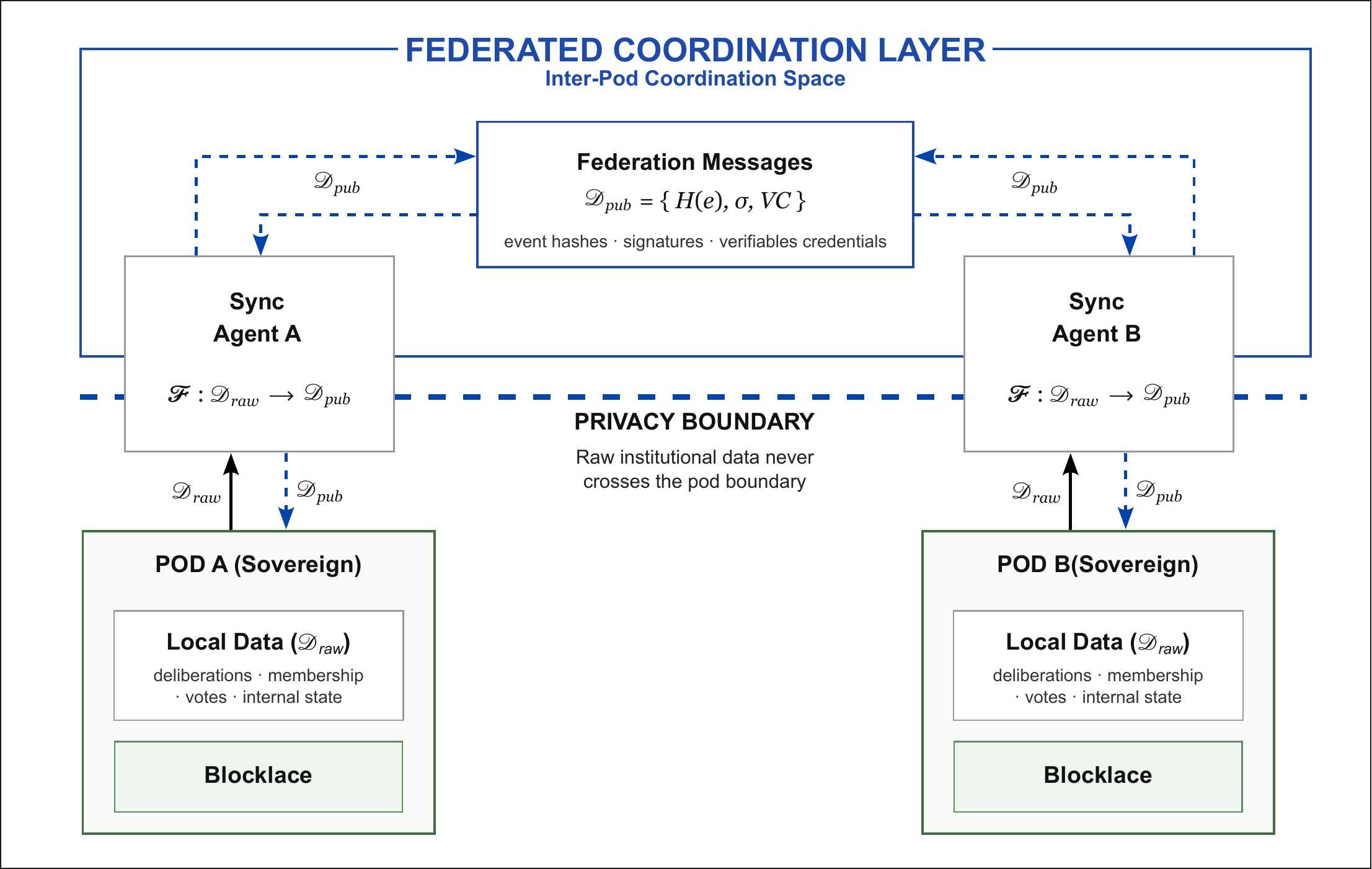}
		\caption{\FSTP{} architecture. Only typed cryptographic artifacts
			($\Dpub$) cross the node boundary through the synchronization
			agent. Raw internal data ($\Draw$) never enters the federation
			layer.}
		\label{fig:architecture}
	\end{figure}

	Each participating institution operates a \textbf{node}: a server
	under its own administrative control, with its internal data store
	encrypted at rest (AES-256-GCM; key derivation via Argon2id) under
	institution-controlled keys. The protocol does not constrain what
	platform runs inside the node; it defines the boundary between what
	stays inside and what may leave. The reference synchronization agent
	is written in Rust. Conforming platform integrations may implement the
	same boundary contract in other languages and stacks; one production
	deployment uses Java~17 with Google Tink for key management and
	BouncyCastle for Ed25519 signing operations. Nodes participate in the
	federation exclusively through the synchronization agent
	(Figure~\ref{fig:architecture}).
	
	The node's information universe is partitioned into two disjoint sets:
	$\Draw$ (raw internal data: membership records, deliberation content,
	identity attributes, operational information) and $\Dpub$ (typed
	cryptographic artifacts: event hashes, digital signatures, typed
	metadata envelopes, and verifiable credentials explicitly authorized
	for propagation). No element of $\Draw$ is derivable from any element
	of $\Dpub$ without breaking the underlying cryptographic primitives,
	so an observer with full access to all federation traffic learns
	nothing about the content of any internal process.
	
	\begin{property}[Synchronization confinement]
		\label{prop:confinement}
		Every message $m$ emitted by the \SA{} toward the federation
		satisfies
		\[
		m \;\cap\; \Draw \;=\; \emptyset.
		\]
	\end{property}

	\begin{property}[Bounded re-emission]
		\label{prop:reemission}
		Let $d \in \Dpub$ be emitted by node $A$ to node $B$ under a usage
		scope $\sigma$ carried in $d$'s typed metadata envelope. For every
		conforming node $B$: if $B$'s synchronization agent emits $d$, or an
		artifact derived from $d$, toward a third node $C$, then either
		$\sigma$ authorizes $C$, or Governance for the underlying assertion
		records a fresh grant extending $\sigma$ to $C$. Absent one of these,
		a conforming $B$ refuses the re-emission. The property binds a
		conforming agent, under the same threat-model boundary as
		Property~\ref{prop:confinement}: it does not bind an operator who
		runs modified code (Section~\ref{sec:threat}).
	\end{property}
	
	\subsection{Threat Model}
	\label{sec:threat}
	
	\FSTP{} addresses four adversarial capabilities. An
	\emph{honest-but-curious federation peer} follows the protocol but
	attempts to infer maximal information from compliant messages. An
	\emph{honest-but-curious infrastructure operator} observes traffic and
	storage metadata. A \emph{passive network interceptor} observes
	ciphertext but cannot break transport-layer encryption. A
	\emph{cross-context correlator} attempts to link a node's contextual
	identities across different federation relationships.
	
	\FSTP{} does not protect against malicious node administrators with
	authorized access to the data store, or against compromise of the
	underlying cryptographic primitives (SHA-256, AES-256-GCM, Ed25519).
	Mitigations available within the architecture for the malicious
	administrator case are discussed in Section~\ref{sec:discussion}.
	
	\subsection{Comparison with Existing Approaches}
	\label{sec:comparison}
	
	\FSTP{}'s design sits at the intersection of four bodies of prior work,
	and its contribution is the integration of their relevant properties
	into a single transport layer that enforces confinement by design.
	
	ActivityPub~\cite{w3c2018activitypub} and Matrix~\cite{matrix2022spec}
	enable decentralized communication but lack structural constraints on
	data emission; confinement rests on operator policy, and both protocols
	allow globally resolvable user identities. The Solid
	project~\cite{berners2016solid} introduces personal data pods that
	decouple storage from application logic, providing structural custody
	of data at rest; however, Solid does not define an inter-node transport
	layer for active federation, so its confinement guarantees do not
	extend to the coordination scenarios \FSTP{} targets.
	Blockchain-based governance~\cite{hassan2021dao} provides structural
	integrity for audit records but requires global visibility of all
	transaction data---the opposite of what privacy-constrained institutions
	need---and makes data erasure structurally impossible, as discussed in
	Section~\ref{sec:erasure}.
	
	The W3C standards for Verifiable Credentials~\cite{w3c2022vc} and
	Decentralized Identifiers~\cite{w3c2022did} provide the identity
	foundation that \FSTP{} builds on. Standard self-sovereign identity
	(\textsc{ssi}) implementations focus on individual-to-service
	relationships and do not address the cross-context correlation problem
	at the transport layer; \FSTP{}'s contextual identity model extends
	their approach to the inter-node boundary. Table~\ref{tab:sota}
	summarizes the key dimensions along which \FSTP{} differs from the
	state of the art.
	
	\begin{table}[h]
		\centering
		\caption{Comparison of federation and data-custody protocols across
			dimensions relevant to privacy-constrained institutions.
			\textit{Structural} denotes a guarantee enforced by the protocol
			itself; \textit{Policy} denotes a guarantee contingent on operator
			behavior; \textit{N/A} denotes that the dimension is not addressed.}
		\label{tab:sota}
		\small
		\begin{tabularx}{\linewidth}{@{} l X X X X X @{}}
			\toprule
			\textbf{Protocol} &
			\textbf{Trust model} &
			\textbf{Identity linkability} &
			\textbf{Sync cost} &
			\textbf{Right to erasure} &
			\textbf{Content confinement} \\
			\midrule
			ActivityPub &
			Policy &
			Global (resolvable) &
			Full fan-out &
			N/A &
			Policy only \\
			\addlinespace[0.5ex]
			Matrix &
			Policy &
			Global (MXID) &
			Room-scoped replication &
			N/A &
			Policy only \\
			\addlinespace[0.5ex]
			Solid &
			Structural (at rest) &
			Per-pod &
			N/A (storage only) &
			Structural (at rest) &
			Structural (at rest) \\
			\addlinespace[0.5ex]
			Blockchain / DAO &
			Structural &
			Global (address) &
			Full replication &
			Incompatible &
			None (transparent) \\
			\addlinespace[0.5ex]
			\FSTP{} (this work) &
			\textbf{Structural} &
			\textbf{Contextual (unlinkable)} &
			\textbf{$O(\Delta)$} &
			\textbf{Structural (dangling pointer)} &
			\textbf{Compile-time} \\
			\bottomrule
		\end{tabularx}
	\end{table}

	\section{Protocol Design}
	\label{sec:design}
	
	\FSTP{} defines three protocol primitives: the synchronization agent,
	the contextual identity model, and the Blocklace event substrate.
	
	\subsection{Synchronization Agent and Data Confinement}
	\label{sec:sa}
	
	The synchronization agent (\SA{}) is the exclusive interface between
	the node's internal data store and the federation network.
	
	\paragraph{Type-system enforcement.}
	\label{sec:typesystem}
	
	Property~\ref{prop:confinement} is enforced by the type system of the
	reference implementation, which is written in Rust. The key mechanism
	is a \emph{closed enumeration} over $\Dpub$: a Rust \texttt{enum}
	whose variants correspond exactly to the message types the \SA{} is
	permitted to emit. Because Rust requires \texttt{match} expressions to
	be exhaustive, and because the internal data types ($\Draw$) are not
	members of this enumeration and are not importable in the message
	module, the compiler statically rejects any code path that attempts to
	construct a federation message from an internal data value.
	
	Formally, let $T_{\text{raw}}$ be the set of Rust types representing
	$\Draw$ (e.g., \texttt{MemberRecord}, \texttt{DeliberationEntry},
	\texttt{IdentityAttribute}), and let $T_{\text{pub}}$ be the closed
	enumeration representing $\Dpub$ (e.g.,
	\texttt{FstpMessage::IdentityEvent},
	\texttt{FstpMessage::VerifiableCredential},
	\texttt{FstpMessage::EventHash},
	\texttt{FstpMessage::FederationControl}). The type system enforces
	$T_{\text{raw}} \cap T_{\text{pub}} = \emptyset$ at the syntactic
	level: any attempt to construct a value of type \texttt{FstpMessage}
	from a value of any type in $T_{\text{raw}}$ is a type error, caught
	before compilation. No runtime check is required or used. Extending
	the output vocabulary---adding a new federation message type---requires
	modifying the \texttt{FstpMessage} enum, a change that is auditable in
	the open-source repository by any federation peer.
	
	The practical significance of this distinction is worth emphasizing.
	A runtime access control check is evaluated at execution time; it can
	fail due to programming errors, be bypassed by a crafted input, or
	produce false positives that block legitimate operation. The Rust
	type-system guarantee is a static property of the program: if the
	program compiles, the confinement property holds for all possible
	executions. The only way to circumvent it deliberately is to modify
	the source of the \texttt{FstpMessage} enum, which is visible and
	auditable.
	
	Listing~\ref{lst:rust} illustrates the enforcement. The function
	\texttt{build\_message} is the only code path that produces a value of
	type \texttt{FstpMessage}; its parameter is \texttt{AuthorizedEvent},
	a type in $T_{\text{pub}}$ produced by the \SA{}'s validation stage.
	A raw internal type such as \texttt{MemberRecord} cannot be passed to
	this function because the compiler rejects the call at the
	type-checking stage, before the program is compiled.
	
	\begin{lstlisting}[language=Rust,
	caption={Simplified Rust fragment illustrating compile-time
	confinement. \texttt{FstpMessage} is the closed output type
	($\Dpub$); \texttt{MemberRecord} is a raw internal type ($\Draw$).
	The commented call below is a type error caught at compile time,
	not at runtime.},
	label=lst:rust]
	// Internal data types (D_raw) -- never cross the node boundary
	struct MemberRecord { name: String, joined: u64 }
	struct DeliberationEntry { content: String, author_id: u64 }
	
	// Authorized staging type produced by the SA's validation stage
	struct AuthorizedEvent {
	event_hash: [u8; 32], class: EventClass,
	timestamp: u64, cii: ContextualId, signature: Signature
	}
	
	// Closed output enumeration (D_pub)
	enum FstpMessage {
	IdentityEvent { cii: ContextualId, pubkey: PublicKey,
	endpoint: Url, timestamp: u64, sig: Signature },
	EventHash     { event: AuthorizedEvent },
	VerifiableCredential { subject_cii: ContextualId,
	credential_type: String, claims: Claims,
	issuer_cii: ContextualId, validity: Validity,
	sig: Signature },
	FederationControl { control_type: ControlType,
	from_cii: ContextualId, to_cii: ContextualId,
	capabilities: Vec<Capability>, sig: Signature },
	}
	
	// The SA's only outbound function accepts D_pub, not D_raw
	fn build_message(event: AuthorizedEvent) -> FstpMessage {
	FstpMessage::EventHash { event }
	}
	
	// This call is a compile-time type error -- confinement is guaranteed:
	// build_message(MemberRecord { name: "Alice".into(), joined: 0 });
	\end{lstlisting}
	
\begin{figure}[h]
		\centering
		\includegraphics[width=\textwidth]{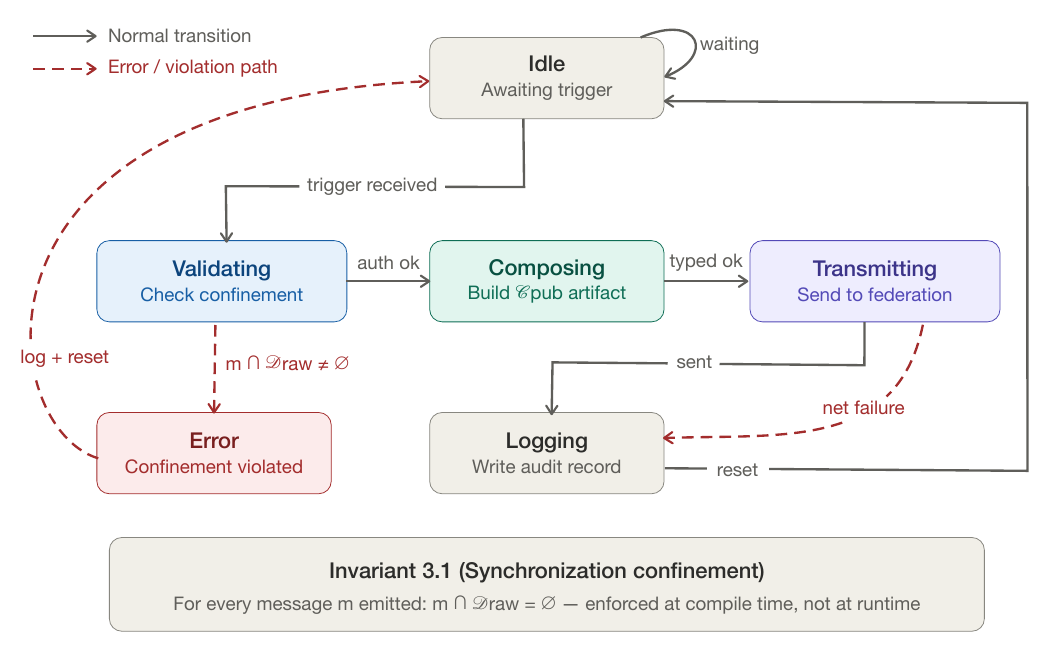}
		\caption{FSTP synchronization agent state machine. Normal
			transitions (solid lines) follow the happy path: \textsc{idle}
			$\to$ \textsc{validating} $\to$ \textsc{composing} $\to$
			\textsc{transmitting} $\to$ \textsc{logging} $\to$ \textsc{idle}.
			Error paths (dashed red) are triggered when
			Property~\ref{prop:confinement} is violated or by a network
			failure; both terminate with an audit log entry before resetting
			to \textsc{idle}.}
		\label{fig:statemachine}
	\end{figure}

	The \SA{} transitions through five states (Figure~\ref{fig:statemachine}):
	\textsc{idle}, \textsc{validating}, \textsc{composing},
	\textsc{transmitting}, and \textsc{logging}. It never enters a state
	in which $\Draw$ values are directly marshalled to network output;
	both confinement violations and network failures terminate with an
	audit log entry before the agent resets to \textsc{idle}.
	
	\paragraph{Base message type contract.}
	
	\FSTP{} defines four base message types (Table~\ref{tab:messages}).
	Deploying networks may extend this set with additional typed objects,
	provided all additions satisfy Property~\ref{prop:confinement}:
	
	\begin{enumerate}[itemsep=2pt,topsep=2pt]
		\item \textbf{Identity events}: contextual identifier, public key,
		federation endpoint.
		\item \textbf{Event hashes}: SHA-256 hash of a certified internal
		event with a typed metadata envelope (event class, timestamp,
		aggregate attributes). No event content leaves the node.
		\item \textbf{Verifiable credentials}: propagated only when the
		credential subject explicitly authorizes it, and scoped to the
		designated recipient.
		\item \textbf{Federation control events}: typed messages that
		establish, update, or terminate federation relationships.
	\end{enumerate}
	
	\begin{table}[h]
		\centering
		\caption{\FSTP{} synchronization agent outbound message types.
			All fields are elements of $\Dpub$; no field may carry a value of
			type $\Draw$, enforcing Property~\ref{prop:confinement} at the
			type-system level.}
		\label{tab:messages}
		\small
		\begin{tabularx}{\linewidth}{@{} l X c @{}}
			\toprule
			\textbf{Type} & \textbf{Key fields} &
			\textbf{Content leaves node?} \\
			\midrule
			\texttt{IdentityEvent} &
			CII, pubkey, endpoint, timestamp, $\sigma$ & No \\
			\addlinespace[1ex]
			\texttt{EventHash} &
			event\_class, timestamp, CII, aggregate\_attrs, $H(e)$, $\sigma$
			& No \\
			\addlinespace[1ex]
			\texttt{VerifiableCredential} &
			subject\_CII, credential\_type, claims, issuer\_CII, validity,
			$\sigma$ & Consent only \\
			\addlinespace[1ex]
			\texttt{FederationControl} &
			control\_type, from\_CII, to\_CII, capabilities, $\sigma$ & No \\
			\bottomrule
		\end{tabularx}
	\end{table}
	
	\paragraph{Local audit log.}
	
	The \SA{} maintains a local audit log visible to the institution's
	administrators. Every outbound message is recorded with its type,
	timestamp, destination, and reference identifier; content is never
	written to the log. This gives a verifiable record of all federation
	activity that does not depend on any external party---a mechanism
	relevant for regulatory compliance when credential wallets are held
	under platform custody.
	
	\subsection{Contextual Identity Model}
	\label{sec:identity}
	
	\paragraph{Global and contextual identifiers.}
	
	Each entity holds a \textbf{Global Identity} (\GII{}): a persistent
	Decentralized Identifier~\cite{w3c2022did} that is the root of its
	cryptographic identity. The reference implementation uses the
	\texttt{did:key} method with Ed25519 keys, producing self-certifying
	identifiers that require no registry and are fully resolvable offline.
	The \GII{} is not used directly in federation
	traffic. For each federation relationship, the \SA{} derives a
	\textbf{Contextual Identity} (\CII{}) from the \GII{} using a
	one-way function parameterized by the relationship context. The \CII{}
	is stable within the relationship, but it cannot be linked to the
	entity's \GII{} or to its \CII{}s in other contexts without the
	entity's cooperation.
	
	\begin{property}[Contextual isolation]
		\label{prop:isolation}
		Let entity $E$ participate in contexts $F_1, \ldots, F_n$ with
		contextual identities $\CII_{F_i} = f(\GII_E, F_i)$, where $f$ is a
		one-way function. Under the one-wayness of $f$, no federation
		participant can determine that $\CII_{F_i}$ and $\CII_{F_j}$
		($i \neq j$) belong to the same entity without $E$'s cooperation.
	\end{property}
	
	\paragraph{Structural prevention of context collapse.}
	
	The contextual identity model is a direct structural response to the
	\emph{context collapse} problem identified in social network
	research~\cite{marwick2011tweet}: the unintended flattening of
	distinct social contexts into a single audience, causing information
	disclosed appropriately in one context to reach actors in a different
	context for whom it was not intended. In existing federated systems, a
	user's global identity is resolvable across all nodes that have
	received any of their messages, which makes context collapse an
	architectural inevitability rather than an exceptional failure mode.
	In \FSTP{}, because each federation relationship derives an
	independent, unlinkable \CII{}, no observer can determine that two
	contextual identities belong to the same entity without that entity's
	active cooperation. This property follows from the one-wayness of the
	derivation function and requires no additional configuration.
	
	\paragraph{Credential wallet and minimum disclosure.}
	
	Each user maintains a wallet of verifiable
	credentials~\cite{w3c2022vc}. When interacting with another node, the
	\SA{} presents only the credential subset relevant to that
	interaction, under the user's \CII{} for that relationship.
	
	\begin{property}[Minimum disclosure]
		\label{prop:mindisclosure}
		A credential presentation under \FSTP{} discloses to the destination
		node exactly the claims in the presented credentials and the \CII{}
		under which they are presented. No additional information about the
		user or their home node is inferable from the presentation message.
	\end{property}
	
	If the destination node accepts the presentation, the user receives
	\textbf{residency}: a contextual identity valid for the duration of
	their active participation in that node, which lapses when active
	membership falls to zero. The user's home node retains the audit record
	of every credential presentation made on their behalf.
	
	\subsection{Blocklace Event Substrate}
	\label{sec:blocklace}
	
	Each node stores its event history as a
	Blocklace~\cite{shapiro2023grassroots,keidar2023cordial}: a
	cryptographic directed acyclic graph of signed blocks that replaces
	linear append-only logs.
	
	\begin{definition}[Blocklace]
		A Blocklace $\Blocklace$ is a set of signed blocks. Each block
		$b \in \Blocklace$ consists of: $\mathrm{payload}(b)$, a non-empty
		event sequence; $\mathrm{parents}(b)$, a set of hash pointers to
		earlier blocks; and $\mathrm{sig}(b)$, a digital signature over
		payload and parents. The frontier $\Fr(\Blocklace)$ is the set of
		blocks with no successors in $\Blocklace$.
	\end{definition}
	
	Unlike a linear log, the Blocklace does not impose a total order.
	Concurrent events from different nodes coexist as parallel branches
	without forced serialization, which accurately represents the temporal
	structure of distributed institutional decisions: fabricating
	precedence among genuinely concurrent decisions can carry legal or
	procedural significance and should be avoided.
	
	\paragraph{Synchronization efficiency.}
	
	The synchronization protocol works by frontier exchange. Each of two
	nodes transmits its frontier $\Fr(\Blocklace)$; blocks are identified
	by cryptographic hash, so each node determines in $O(|\Fr|)$ time
	which frontier blocks are locally absent, then requests those blocks
	and all unobserved ancestors. Because every block in the frontier of
	one node that is absent from the other must be transmitted, and no
	block that both nodes already hold need be sent, the set of blocks
	exchanged is exactly the symmetric difference $\Delta$ between the two
	Blocklaces. Synchronization cost is therefore $O(\Delta)$, independent
	of total Blocklace size: a node that reconnects after an absence
	exchanges only the events produced during that absence, not the full
	history.
	
	\begin{remark}
		\label{remark:odelta}
		The $O(\Delta)$ bound describes the number of blocks transmitted as
		a function of the symmetric difference between node states, and is a
		consequence of the frontier-exchange algorithm. It characterizes
		protocol efficiency in terms of information exchanged, not in terms
		of wall-clock latency or bandwidth under specific network conditions.
		Empirical validation of this bound for the reference implementation
		is reported in Section~\ref{sec:benchmarks}.
	\end{remark}
	
	\paragraph{Tamper evidence.}
	
	Modification of any block changes its hash, invalidating all hash
	pointers in descendant blocks. A node holding the hash of a later
	block can therefore detect retroactive modification of any ancestor,
	and a verifier can confirm the integrity of any event in the history
	by checking the chain from any later block it holds.
	
	\paragraph{Erasure compatibility and the dangling pointer mechanism.}
	\label{sec:erasure}
	
	Events are stored in the node's data store; the Blocklace holds hash
	\emph{pointers} to events, not the events themselves. This structural
	separation between content and structure resolves an incompatibility
	that is architectural in linear ledger designs.
	
	In a conventional blockchain, each block contains the data of the
	transactions it records. Deleting data from a block changes its hash,
	breaking the hash pointer in the following block and cascading through
	the entire chain. Because integrity depends on all blocks being
	present and unmodified, deletion is structurally incompatible with
	ledger integrity: the two obligations cannot be satisfied
	simultaneously. Legal frameworks that impose both an immutable audit
	trail requirement and a right to erasure---such as the
	GDPR~\cite{gdpr2016} or Mexico's
	\textsc{lfpdppp}~\cite{lfpdppp2010}---therefore cannot both be
	satisfied by a linear blockchain architecture, regardless of how that
	architecture is configured. This is a structural limitation of linear
	ledgers, not a deployment problem.
	
	The Blocklace resolves this by decoupling content from structure. When
	an event is deleted from the data store to fulfill an erasure
	obligation, the corresponding block becomes a \textbf{dangling
		pointer}:
	
	\begin{itemize}[itemsep=2pt]
		\item Block $b_2$ remains in the Blocklace with its hash $H(e_2)$,
		timestamp, and signature intact.
		\item Content $e_2$ is deleted from the data store and is
		irrecoverably absent.
		\item Blocks $b_3$ and $b_4$ reference $b_2$ through $H(b_2)$---the
		hash of the \emph{block}, not of its content. Because $b_2$ itself
		is unmodified, $H(b_2)$ remains valid.
		\item Tamper-evidence is preserved for all descendant blocks: a
		verifier holding $b_4$ can confirm the integrity of the full
		chain, but cannot recover $e_2$.
	\end{itemize}
	
	The block is a pointer, not a container; deleting the pointed-to
	content leaves the pointer intact. The chain does not break---it
	simply points to something that is no longer present. This design
	allows an institution to satisfy a data erasure request by deleting
	event content from its data store while preserving the structural
	record that the event occurred, when it occurred, and that it has not
	been retroactively altered.
	
	\begin{property}[Erasure without integrity loss]
		\label{prop:erasure}
		Let $b \in \Blocklace$ be the block whose payload references event
		$e$. Deletion of $e$ from the data store produces a dangling pointer
		at $b$ while leaving $H(b)$ unchanged. For all blocks $b'$ such that
		$b \leq_{\Blocklace} b'$, tamper-evidence is preserved: $H(b)$
		remains a valid ancestor reference in the hash chain.
	\end{property}
	
	\begin{figure}[h]
		\centering
		\includegraphics[width=\textwidth]{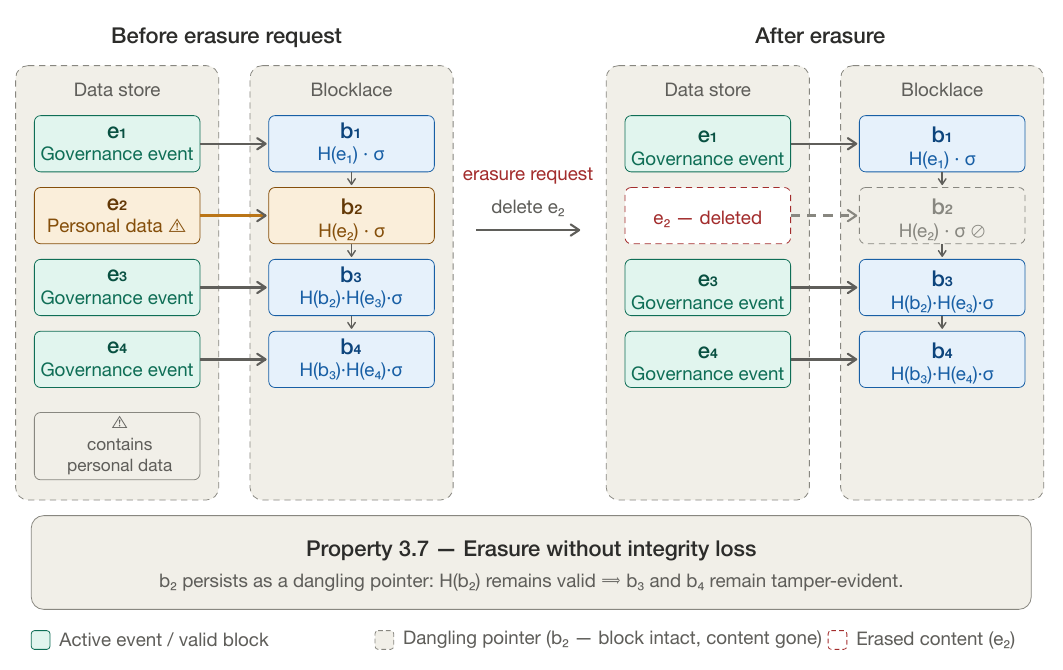}
		\caption{Blocklace erasure mechanism. \textit{Left}: four blocks
			$b_1$--$b_4$ reference events $e_1$--$e_4$ in the data store;
			$e_2$ contains data subject to an erasure obligation.
			\textit{Right}: after erasure, $e_2$ is deleted; $b_2$ becomes a
			dangling pointer---$H(b_2)$ remains unchanged, so $b_3$ and $b_4$
			remain tamper-evident. Erasure and audit trail integrity are
			satisfied simultaneously.}
		\label{fig:dangling}
	\end{figure}
	
	Whether a particular deployment thereby satisfies all applicable
	requirements---GDPR Article~17, Mexico's \textsc{lfpdppp}, or
	analogous frameworks in other jurisdictions---depends on
	jurisdictional analysis beyond the scope of this paper. The
	contribution here is architectural: the structural obstacle present
	in linear ledgers is removed.

	\section{Security and Privacy Properties}
	\label{sec:security}
	
	\subsection{Security Properties}
	
	Based on the threat model of Section~\ref{sec:threat}, \FSTP{}
	provides the following guarantees:
	
	\begin{itemize}[itemsep=4pt]
		
		\item \textbf{Data confidentiality.} Node contents are encrypted at
		rest under institution-controlled keys. Access without
		administrator credentials yields only ciphertext.
		
		\item \textbf{Synchronization confinement.}
		Property~\ref{prop:confinement} is enforced by the Rust type system
		as a compile-time structural guarantee. The closed output
		enumeration means confinement cannot be accidentally violated;
		deliberate circumvention requires modifying the open-source agent
		code.
		
		\item \textbf{Event integrity.} The Blocklace hash structure provides
		tamper-evidence cascading to all descendant blocks. Federation
		participants cannot fabricate event hashes without the issuing
		node's private signing key.
		
		\item \textbf{Credential unforgeability.} Verifiable credentials are
		signed by the issuing node's private key and verifiable by any
		relying party through DID resolution, without contacting the
		issuer.
		
		\item \textbf{Contextual isolation.} Property~\ref{prop:isolation}:
		contextual identities across different federation relationships are
		unlinkable without the entity's cooperation.
		
		\item \textbf{Minimum disclosure.} Property~\ref{prop:mindisclosure}:
		a credential presentation discloses exactly the presented claims and
		the contextual identity---nothing more.
		
	\end{itemize}
	
	\subsection{Protocol-Level Privacy Audit}
	\label{sec:privacyaudit}
	
	The \textbf{information surface} of a protocol is the set of facts
	that an honest-but-curious observer of conforming traffic can infer
	from the messages exchanged. The following audit characterizes this
	surface for \FSTP{} under the threat model of
	Section~\ref{sec:threat}.
	
	An honest-but-curious observer of conforming \FSTP{} traffic
	\textbf{learns}: that a contextual identity is active in a specific
	federation relationship; that events of given classes occurred at
	given times with given aggregate characteristics; which contextual
	identities are linked within observable contexts; and that specific
	credential types were presented to specific destinations where
	authorized.
	
	The same observer \textbf{cannot learn}: the content of any internal
	event; individual participant identities in any process; individual
	positions within any deliberation; the mapping between contextual and
	global identities; the presenting user's wallet contents beyond the
	presented subset; or the node's participation in other federation
	relationships.
	
	\paragraph{Residual: traffic metadata.}
	
	An observer may infer governance activity rhythms from event timing
	and frequency. This risk is real but bounded: the observer learns
	\emph{that} governance activity occurred and \emph{when}, but not
	\emph{what} was decided or \emph{who} participated---the content of
	the process and the identities involved remain inaccessible. For
	deployments where even activity rhythms are sensitive, randomized
	announcement delays are compatible with \FSTP{} because the Blocklace
	relies on partial ordering rather than wall-clock synchronization, so
	delays do not affect correctness. Formal quantification of information
	leakage from traffic patterns under a specified adversary model is
	left for future work.
	
	\subsection{Regulatory Compatibility}
	
	Property~\ref{prop:erasure} is designed to enable simultaneous
	satisfaction of audit trail integrity requirements and data erasure
	obligations. The dangling pointer mechanism decouples content deletion
	from hash-chain integrity: content deletion addresses erasure
	requirements; the persisting hash record satisfies integrity
	requirements. This resolution applies regardless of the specific legal
	framework in force, because it rests on the structural properties of
	the Blocklace rather than on any jurisdiction-specific configuration.
	As noted in Section~\ref{sec:erasure}, whether a particular deployment
	achieves full compliance with GDPR Article~17, Mexico's
	\textsc{lfpdppp}, or analogous frameworks in other jurisdictions
	requires jurisdictional analysis beyond the scope of this paper.
	
	Property~\ref{prop:mindisclosure} aligns with data minimization
	principles present in contemporary data protection frameworks across
	jurisdictions. When wallets are held under platform custody, the audit
	log of every credential presentation provides the accountability
	record that such frameworks typically require.

	\section{Coordination Scenarios}
	\label{sec:scenarios}
	
	The following scenarios demonstrate how the three protocol primitives
	compose under different coordination requirements. They are
	deliberately generic---\FSTP{} is a transport protocol and the
	institutions that deploy it vary in sector, size, and legal context.
	What they share is the need to coordinate across node boundaries while
	retaining exclusive custody of internal data.
	
	\subsection{Case Study: Federated Decision Across Sovereign Nodes}
	\label{sec:scenario:b}
	
	This case study develops a complete protocol trace for a
	three-institution governance scenario. It is chosen because it
	exercises all three protocol primitives simultaneously and
	illustrates the governance accountability question: how can an
	external auditor verify the legitimacy of a collective outcome without
	accessing the deliberative record of any participating institution?
	
	\paragraph{Structure.}
	
	Three institutions---$A$, $B$, and an aggregation node $C$---must
	jointly produce a certified collective decision. Institutions $A$ and
	$B$ each conduct an internal deliberation and vote independently. The
	outcome of each institution's process must remain inaccessible to the
	other institution and to $C$; what the network requires is verifiable
	proof that each institution completed a legitimate process and
	produced a certified outcome.
	
	\paragraph{Threat context.}
	
	Nodes $A$ and $B$ are honest-but-curious peers, each attempting to
	infer the other's internal process from federation traffic. Node $C$
	is an honest-but-curious aggregation node attempting to infer
	participation patterns from the event hashes it receives. A subsequent
	external audit requires the auditor to verify the legitimacy of the
	collective outcome without receiving the deliberative record of any
	institution.
	
	\paragraph{Protocol trace.}
	
	\begin{enumerate}[itemsep=3pt,topsep=4pt]
		
		\item \textbf{Federation link establishment.} Nodes $A$, $B$, and
		$C$ exchange \texttt{FederationControl} messages to establish
		mutually authenticated channels. Each node publishes its CII and
		public key via \texttt{IdentityEvent}. No internal data is
		exchanged in this phase.
		
		\item \textbf{Internal deliberation (nodes $A$ and $B$, in parallel).}
		Each institution conducts its deliberation entirely within its own
		node. The deliberation content---agenda, contributions, votes---is
		of type $\Draw$ and does not leave either node at any stage.
		
		\item \textbf{Event hash emission.} Upon certification of its
		internal decision, each \SA{} constructs an \texttt{EventHash}
		envelope:
		\[
		\langle\, H(e_i),\ \texttt{class:decision},\
		\texttt{ts}_i,\ \sigma_i \,\rangle,
		\]
		where $H(e_i)$ is the SHA-256 hash of the certified event and
		$\sigma_i$ is computed over the envelope using the node's Ed25519
		signing key. No event content appears in the envelope. The two
		envelopes are transmitted to $C$ over mutually authenticated
		\FSTP{} channels.
		
		\item \textbf{Aggregation at node $C$.} Node $C$ verifies the
		signatures $\sigma_A$ and $\sigma_B$ and the typed metadata (event
		class \texttt{decision}, timestamps within the protocol window). It
		does not---and does not need to---verify the content of either
		institution's deliberation. Node $C$ applies the network's
		aggregation rule to $(H(e_A), H(e_B))$ and records the collective
		outcome in its Blocklace. The two input hashes appear as parallel
		branches, preserving the genuine concurrency of the institutional
		decisions without imposing an artificial ordering.
		
		\item \textbf{External audit.} An external authority requests proof
		of the legitimacy of institution $A$'s contribution. Institution
		$A$ discloses its full internal deliberation record directly to
		the authority. The authority computes $H(e_A)$ from the disclosed
		record and compares it against the \texttt{EventHash} already
		present in node $C$'s Blocklace. A match provides cryptographic
		proof that the disclosed record is authentic, that it was produced
		before the collective outcome was recorded, and that it has not
		been retroactively altered---without requiring the authority to
		contact node $C$, node $B$, or any other party. Institution $B$'s
		deliberation is not disclosed in this process.
		
	\end{enumerate}
	
	This trace shows the central accountability property of \FSTP{}. The
	auditor can verify the legitimacy of the collective outcome with the
	same cryptographic confidence as if they had observed the deliberation
	in real time, without accessing the deliberation during the governance
	process itself. The Blocklace record is the independently verifiable
	anchor; the institution's local record is the evidence; the hash is
	the link between them. A centralized audit architecture cannot provide
	this property because the auditor must trust that a central operator
	has not modified the record; in \FSTP{}, the auditor trusts the hash
	function and the public key infrastructure, both of which are
	verifiable without relying on any party's honesty.
	
	Throughout the trace, Property~\ref{prop:confinement} holds: the only
	values crossing any node boundary are elements of $\Dpub$. The
	deliberation content $e_A$ and $e_B$ are elements of $\Draw$ and
	remain at their respective nodes at all times.
	
	\subsection{Additional Coordination Patterns}
	\label{sec:scenario:others}
	
	\paragraph{User crossing a node boundary.}
	
	A user on a shared platform node joins a group on a privately operated
	node. The \SA{} derives a \CII{} for the relationship and presents
	only the wallet credentials relevant to the group's access policy. The
	destination node grants residency under the user's \CII{}; interaction
	proceeds through the \SA{}. Neither node's data store is exposed to
	the other. When the user leaves, the destination node retains only the
	\CII{} and the presented credentials---it learns nothing about the
	user's home node, global identity, or other federation relationships.
	The home node retains the audit record of every credential
	presentation made on the user's behalf, satisfying the accountability
	requirement of Property~\ref{prop:mindisclosure} in its bilateral
	form.
	
	\paragraph{Verifiable credential network.}
	
	A certification authority issues signed credentials to member
	organizations. Holders present credentials to relying parties under
	their contextual identity for that relationship. The relying party
	verifies by DID resolution---without contacting the issuer---which
	means the issuer is not informed of individual presentations and
	cannot construct a usage profile across relying parties. Revocation
	propagates through the federation via DID document update. The
	holder's wallet log records every presentation. This arrangement
	applies Property~\ref{prop:mindisclosure} in its hub-and-spoke form:
	the issuing node certifies once, relying parties verify independently,
	and the holder controls disclosure.
	
	\subsection{Summary}
	
	Three structurally distinct patterns---bilateral user-to-node,
	multilateral inter-institutional, hub-and-spoke credential
	network---are handled by the same three protocol primitives without
	modification. The same \SA{} confinement property, the same contextual
	identity model, and the same Blocklace substrate compose differently
	for each pattern. This composability is the primary empirical claim of
	the scenarios: \FSTP{} is general with respect to the coordination
	structures that privacy-constrained institutions encounter in practice.

	\section{Deployment Considerations}
	\label{sec:discussion}
	
	\paragraph{Trust localization.}
	
	\FSTP{} does not eliminate trust requirements; it localizes them to
	the institutional administrative domain. The risk of data exfiltration
	by a local administrator is intrinsic to any sovereign custody model
	and is not resolved by the protocol alone. Four mitigations are
	available within the architecture: the open-source \SA{} code is
	auditable before deployment; the local audit log makes all outbound
	federation activity inspectable by the institution's own
	administrators; the \SA{} software version can be declared in the
	node's DID document, allowing federation peers to verify a conforming
	implementation before establishing a trust relationship; and the
	node's institutional DID can be protected by an $M$-of-$N$ key
	recovery scheme in which a quorum of designated administrators must
	cooperate to authorize DID rotation, preventing a single administrator
	from unilaterally changing the node's cryptographic identity. Full
	mitigation of the malicious administrator threat would require trusted
	execution environments (TEEs), which are architecturally compatible
	with \FSTP{} but outside the current specification.

	\paragraph{Reference deployment modes.}
	
	The reference synchronization agent is a lightweight process that
	implements the \FSTP{} boundary. Deployments may adopt it in three
	common modes: \emph{embedded} (bundled with a local administrator
	application, requiring no additional infrastructure); \emph{server
	service} (running as a system service on organization-owned hardware,
	enabling continuous synchronization independent of whether a desktop
	client is open); and \emph{container} (packaged as an OCI image for
	organizations with existing container infrastructure). In all three
	modes the agent is the exclusive process with outbound network access;
	its traffic log is exposed locally and is auditable by the
	organization's own technical staff and by external auditors without
	relying on any federation operator or platform vendor. The data store
	in all modes is encrypted at rest with AES-256-GCM; the encryption key
	is derived from the administrator's passphrase via Argon2id. The
	private key of the institutional DID is never transmitted; all signing
	operations requiring it are performed locally. One production
	integration ships this agent alongside Velyzor, a governance platform
	for institutions with demanding confidentiality requirements; the
	protocol boundary is identical regardless of the enclosing application.
	
	\paragraph{Extensibility.}
	
	The base message type contract is extensible. Deploying networks may
	define additional typed objects---such as domain-specific governance
	primitives---provided all additions satisfy
	Property~\ref{prop:confinement}. The protocol defines the confinement
	boundary and the extensibility contract; it does not fix the
	application-layer vocabulary.
	
	\paragraph{Scalability and hierarchical federation topologies.}
	
	The $O(\Delta)$ per-pair synchronization cost assumes a flat topology
	in which every node synchronizes directly with every peer.
	Observations of existing federated networks suggest that activity in
	practice concentrates on a small number of high-traffic nodes, a
	pattern consistent with power-law degree distributions documented in
	the Fediverse~\cite{la2021understanding}. In such environments, a
	small set of nodes accumulates large $|\Delta|$ values relative to the
	rest, creating a load imbalance that flat synchronization does not
	address.
	
	\FSTP{}'s Blocklace substrate is natively compatible with hierarchical
	federation topologies, in which each node synchronizes directly with a
	designated subset of peers. In a two-level hierarchy, leaf nodes
	synchronize only with their assigned aggregation node, incurring
	$O(\Delta_{\text{local}})$ cost where $\Delta_{\text{local}}$ reflects
	only events relevant to their subtree. Aggregation nodes bear the
	cross-subtree synchronization cost and can be provisioned accordingly.
	Property~\ref{prop:confinement} applies at every level of the
	hierarchy: no raw internal data crosses any node boundary regardless
	of hierarchy depth. A practical consequence is that institutions with
	intermittent connectivity or limited server capacity can join the
	federation as leaf nodes, synchronizing with a single peer on their
	own schedule. Formal analysis of optimal hierarchy construction and
	empirical benchmarking of synchronization performance under realistic
	network conditions are left for future work.
	
	\paragraph{Structural compliance.}
	
	Data sovereignty in \FSTP{} is not a configuration option but a
	structural consequence of deployment. An institution that deploys a
	conforming node for local operational reasons---compliance
	documentation, credential issuance, internal governance records---
	acquires data sovereignty as an architectural byproduct. The protocol
	does not require an explicit commitment to decentralization as a
	precondition for benefiting from the confinement guarantee.
	
	\paragraph{Open-source release and license.}
	
	The \FSTP{} specification and the reference implementation of the
	synchronization agent are released as open infrastructure. The
	reference implementation is published under the Apache~2.0 License,
	which permits free use, modification, and distribution including in
	commercial and proprietary products, subject to attribution. The
	protocol specification is published under CC~BY~4.0.
	
	The choice of Apache~2.0 is not incidental to the security argument.
	The paper establishes that deliberate circumvention of
	Property~\ref{prop:confinement} requires a modification to the
	\texttt{FstpMessage} enum that is \emph{auditable in the open-source
		repository by any federation peer before establishing a trust
		relationship}. That argument requires that any peer actually has the
	legal right to inspect, compile, and verify the code at any time---not
	merely to read it. Apache~2.0 provides this right unconditionally. A
	source-available license with a commercial protection period, such as
	the Business Source License, would leave the security claim incomplete
	for the duration of that period, because the right to verify would be
	restricted to reading rather than to the full audit cycle of inspect,
	compile, and run. Institutions that integrate \FSTP{} into proprietary
	governance platforms retain full freedom to do so; Apache~2.0 imposes
	no copyleft obligations.

	\section{Empirical Synchronization Performance}
	\label{sec:benchmarks}
	
	Remark~\ref{remark:odelta} characterizes the $O(\Delta)$ cost of the
	frontier-exchange algorithm in terms of blocks transmitted. This
	section reports controlled microbenchmarks that validate this bound
	empirically for the reference implementation and quantify the
	throughput of both the emission and reception sides of a
	synchronization round.
	
	\subsection{Methodology}
	
	Benchmarks were executed with Criterion.rs (v0.5), which collects 100
	samples per configuration, applies outlier filtering, and reports
	two-sided confidence intervals at the 95\% level. All measurements
	were taken on a single machine; inter-node network latency is
	explicitly out of scope and identified as a direction for future work
	(Section~\ref{sec:discussion}).
	
	Each experiment instantiates a pair of \texttt{InMemoryBlocklace}
	nodes $(\mathcal{B}_A, \mathcal{B}_B)$ sharing a common history of
	$N$ blocks, after which $\mathcal{B}_A$ diverges by appending
	$\Delta$ additional blocks that $\mathcal{B}_B$ does not hold. The
	two experimental parameters are varied independently:
	
	\begin{itemize}[itemsep=2pt]
		\item \textbf{Experiment~1} fixes $N = 1{,}000$ and varies
		$\Delta \in \{10, 50, 100, 200, 500\}$, measuring the cost of
		\texttt{sync\_delta} as a function of the symmetric difference
		alone.
		\item \textbf{Experiment~2} fixes $\Delta = 50$ and varies
		$N \in \{100, 500, 1{,}000, 5{,}000\}$, measuring whether the
		cost of \texttt{sync\_delta} depends on the shared history size.
		\item \textbf{Experiment~3} fixes $N = 1{,}000$ and varies
		$\Delta \in \{10, 50, 100, 200\}$, measuring the cost of
		\texttt{merge\_blocks} (block reception and integration) using
		\texttt{iter\_batched} to isolate setup from measurement.
	\end{itemize}
	
	Setup cost (blocklace construction) is excluded from all timing
	loops. Experiment~3 uses \texttt{iter\_batched} so that allocating a
	fresh receiver instance is not counted toward \texttt{merge\_blocks}
	time.
	
	\subsection{Results}
	
	\paragraph{Experiment~1: emission cost scales linearly with $\Delta$.}
	
	Table~\ref{tab:bench1} reports median wall-clock time and throughput
	for \texttt{sync\_delta} as $\Delta$ grows from 10 to 500 blocks,
	with $N = 1{,}000$ held constant.
	
	\begin{table}[h]
		\centering
		\caption{Emission cost (\texttt{sync\_delta}) as a function of
			$\Delta$, with shared history $N = 1{,}000$ blocks fixed.
			Throughput is reported in elements per second; confidence
			intervals are at the 95\% level.}
		\label{tab:bench1}
		\small
		\begin{tabular}{@{} r r r @{}}
			\toprule
			$\Delta$ (blocks) & Median time & Throughput \\
			\midrule
			10  & \phantom{0}5.8~$\mu$s  & 1.72~Melem/s \\
			50  & 21.2~$\mu$s            & 2.36~Melem/s \\
			100 & 40.6~$\mu$s            & 2.46~Melem/s \\
			200 & 79.0~$\mu$s            & 2.53~Melem/s \\
			500 & 221.3~$\mu$s           & 2.26~Melem/s \\
			\bottomrule
		\end{tabular}
	\end{table}
	
	Wall-clock time scales proportionally with $\Delta$: a tenfold
	increase in $\Delta$ (from 50 to 500 blocks) produces a tenfold
	increase in elapsed time ($21.2~\mu\text{s}$ to $221.3~\mu\text{s}$).
	Throughput remains stable at approximately $2.3$~Melem/s across all
	configurations, confirming that the per-block cost is constant and
	that the algorithm is $O(\Delta)$ in practice, not merely in the
	asymptotic analysis.
	
	\paragraph{Experiment~2: emission cost is independent of $N$.}
	
	Table~\ref{tab:bench2} reports \texttt{sync\_delta} time as $N$
	grows from 100 to 5{,}000 blocks with $\Delta = 50$ fixed.
	
	\begin{table}[h]
		\centering
		\caption{Emission cost (\texttt{sync\_delta}) as a function of
			shared history size $N$, with $\Delta = 50$ blocks fixed.}
		\label{tab:bench2}
		\small
		\begin{tabular}{@{} r r r @{}}
			\toprule
			$N$ (shared blocks) & Median time & Throughput \\
			\midrule
			100   & 20.5~$\mu$s & 2.44~Melem/s \\
			500   & 20.6~$\mu$s & 2.42~Melem/s \\
			1{,}000 & 21.9~$\mu$s & 2.28~Melem/s \\
			5{,}000 & 21.2~$\mu$s & 2.36~Melem/s \\
			\bottomrule
		\end{tabular}
	\end{table}
	
	Median time is statistically indistinguishable across a 50-fold
	increase in $N$ (all pairwise comparisons yield $p > 0.05$ under
	Criterion's Welch $t$-test). This confirms the central efficiency
	claim of Section~\ref{sec:blocklace}: the set of blocks that must be
	transmitted is determined by $\Delta$, the symmetric difference
	between node states, and the size of the shared history imposes no
	additional cost on the emitting side.
	
	\paragraph{Experiment~3: reception cost after implementation fix.}
	
	The initial \texttt{merge\_blocks} implementation rebuilt the entire
	frontier cache after each merge, incurring $O(N)$ cost regardless of
	$\Delta$. This was identified as an implementation deficiency relative
	to the $O(\Delta)$ protocol claim: at $\Delta = 10$ over a shared
	history of $N = 1{,}000$ blocks, \texttt{merge\_blocks} took
	${\approx}5.2$~ms---three orders of magnitude slower than the
	corresponding \texttt{sync\_delta} call at $5.5~\mu$s. The frontier
	cache is now updated incrementally: only the parents of newly merged
	blocks are removed from the frontier set, and only the new block
	hashes are inserted, preserving the $O(\Delta)$ invariant on the
	receiving side. Table~\ref{tab:bench3} reports post-fix results.
	
	\begin{table}[h]
		\centering
		\caption{Reception cost (\texttt{merge\_blocks}) as a function of
			$\Delta$, with $N = 1{,}000$ fixed. Timing excludes receiver
			setup (\texttt{iter\_batched}). The previous implementation
			incurred ${\approx}5$--$6$~ms for all $\Delta$ due to full
			frontier cache reconstruction; the incremental fix reduces
			cost by 91--98\%.}
		\label{tab:bench3}
		\small
		\begin{tabular}{@{} r r r r @{}}
			\toprule
			$\Delta$ (blocks) & Median time & Throughput & Reduction vs.\ prior \\
			\midrule
			10  & \phantom{0}77~$\mu$s  & 129~Kelem/s  & ${\approx}98\%$ \\
			50  & 169~$\mu$s            & 296~Kelem/s  & ${\approx}97\%$ \\
			100 & 284~$\mu$s            & 352~Kelem/s  & ${\approx}95\%$ \\
			200 & 526~$\mu$s            & 380~Kelem/s  & ${\approx}91\%$ \\
			\bottomrule
		\end{tabular}
	\end{table}
	
	Reception time now grows proportionally with $\Delta$. The remaining
	gap between emission and reception throughput (${\approx}2.3$~Melem/s
	vs.\ ${\approx}290$--$380$~Kelem/s) reflects legitimate additional
	work on the receiving side: each block undergoes cryptographic hash
	verification (\texttt{compute\_hash}) and insertion into the backing
	\texttt{HashMap}, operations absent from the read-only
	\texttt{sync\_delta} path. Both sides of a synchronization round are
	therefore $O(\Delta)$ in the reference implementation.
	
	\subsection{Summary}
	
	Three experimental results jointly support the $O(\Delta)$ claim of
	Section~\ref{sec:blocklace}:
	
	\begin{enumerate}[itemsep=2pt]
		\item Emission time scales linearly with $\Delta$ at constant
		throughput (${\approx}2.3$~Melem/s), confirming $O(\Delta)$ on
		the sending side.
		\item Emission time is insensitive to $N$ over a 50-fold range,
		confirming that shared history imposes no overhead on the
		frontier-exchange computation.
		\item Reception time, after an incremental frontier-cache update,
		also scales linearly with $\Delta$, with throughput growing from
		129 to 380~Kelem/s as block-level hash verification dominates at
		larger $\Delta$.
	\end{enumerate}

	\begin{figure}[h]
		\centering
		\includegraphics[width=0.75\textwidth]{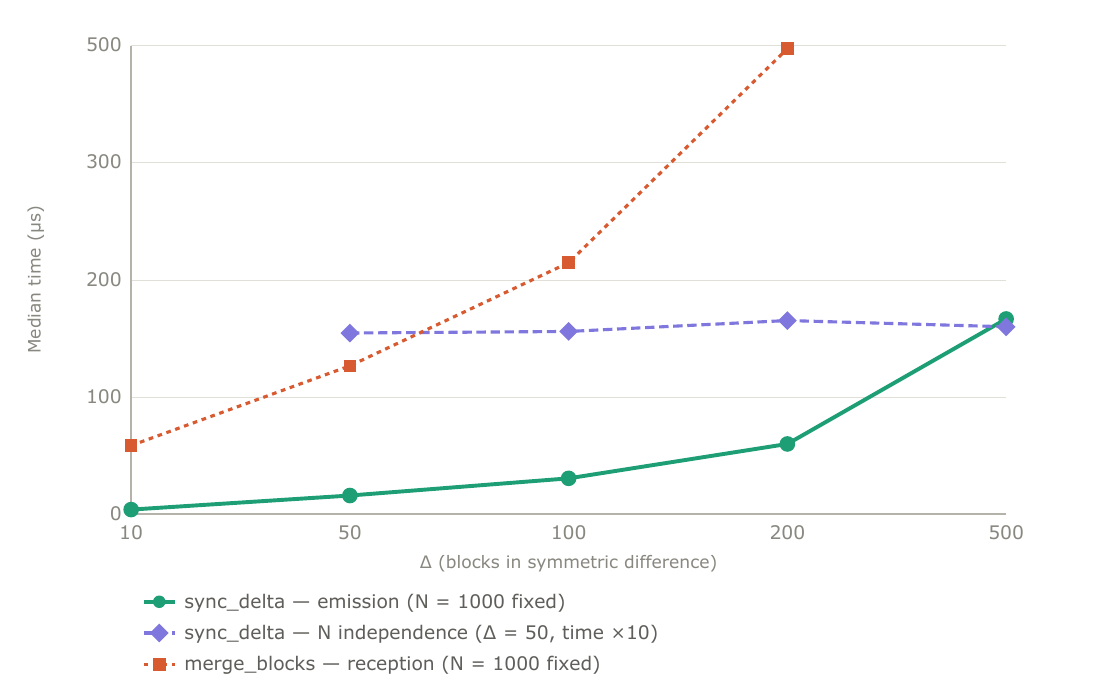}
		\caption{Empirical validation of the $O(\Delta)$ synchronization claim. Emission ($sync\_delta$) scales with the symmetric difference $\Delta$ and remains insensitive to shared history size $N$; reception ($merge\_blocks$) also scales linearly with $\Delta$ after the incremental frontier-cache fix.}
		\label{fig:benchmarking}
	\end{figure}
	
	These results characterize the reference implementation under
	synthetic load with sequential block histories. Benchmarking under
	realistic conditions---concurrent writers, branching DAG topologies,
	and intermittent connectivity---is left for future work.

	\section{Conclusions}
	\label{sec:conclusions}
	
	This paper has presented \FSTP{}, a transport and synchronization
	boundary protocol for federated networks in which nodes have
	heterogeneous privacy requirements.
	
	Three protocol primitives form a coherent confinement architecture.
	The \textbf{synchronization agent} enforces
	Property~\ref{prop:confinement}---no raw internal data in any
	federation message---as a compile-time structural guarantee implemented
	through Rust's exhaustive \texttt{enum} pattern matching, and
	Property~\ref{prop:reemission}---a conforming agent re-emits a received
	artifact only when the usage scope or a fresh governance grant
	authorizes the next recipient. The
	\textbf{contextual identity model} provides unlinkable per-relationship
	identities and minimum-disclosure credential presentation
	(Properties~\ref{prop:isolation} and~\ref{prop:mindisclosure}),
	structurally preventing cross-context correlation and context collapse.
	The \textbf{Blocklace substrate} delivers tamper-evident, partially
	ordered, erasure-compatible event logging whose synchronization cost is
	proportional to the difference between node states
	(Property~\ref{prop:erasure}).
	
	The comparison in Section~\ref{sec:comparison} establishes that
	existing federation protocols address subsets of these properties but
	none addresses all four simultaneously at the protocol level. The
	privacy audit of Section~\ref{sec:privacyaudit} characterizes the
	information surface: an honest-but-curious observer learns that events
	occurred, when, and with what aggregate characteristics, but cannot
	learn event content, individual participant identities, or
	cross-context identity links. The case study of
	Section~\ref{sec:scenario:b} shows that an external auditor can verify
	the legitimacy of a collective governance outcome with cryptographic
	confidence, without accessing the deliberative record of any
	participating institution. This asymmetry is the protocol's central
	property: \textbf{proof without exposure}.
	
	\FSTP{} specifies a transport boundary that conforming platforms may
	adopt independently of any single vendor. The protocol specification
	and reference implementation of the synchronization agent are
	available as open infrastructure under Apache~2.0. Future work includes empirical benchmarking under realistic
	network conditions with intermittent connectivity, formal quantification
	of traffic-metadata leakage under a specified adversary model, and
	analysis of optimal hierarchical federation topology construction.

	\paragraph{Data availability.}
	The \FSTP{} reference implementation (\texttt{fstp-core}, reference
	synchronization agent) and the protocol specification are published
	under Apache~2.0 and CC~BY~4.0, respectively. This paper is available
	at \url{https://arxiv.org/abs/2607.00213}; source and benchmarks are
	at \url{https://github.com/rsotoc/fstp}. Benchmarks in
	Section~\ref{sec:benchmarks} can be reproduced with
	\texttt{cargo bench -p fstp-core} on release tag \texttt{v0.1.0}.

	\bibliographystyle{plainnat}

\end{document}